\documentclass[aps, twocolumn, preprintnumbers]{revtex4}

\usepackage{graphicx}
\usepackage{dcolumn}
\usepackage{bm}
\usepackage{color}

\begin{document}


\title{Gigahertz quantum key distribution with InGaAs avalanche photodiodes}

\author{Z. L. Yuan}
\email{zhiliang.yuan@crl.toshiba.co.uk}

\author{A. R. Dixon}
 \altaffiliation[Also at ]{Cavendish Laboratory, University of Cambridge, J. J. Thomson
Avenue, Cambridge CB3 0HE, UK.}

\author {J. F. Dynes}

\author {A. W. Sharpe}

\author {A. J. Shields}

\affiliation{Toshiba Research Europe Ltd, Cambridge Research
Laboratory, 208 Cambridge Science Park, Milton Road, Cambridge, CB4
0GZ, UK }

\date{\today}

\begin{abstract}

We report a demonstration of quantum key distribution (QKD) at GHz clock rates with InGaAs avalanche photodiodes (APDs) operating in a self-differencing mode.  Such a mode of operation allows detection of extremely weak avalanches so that the detector afterpulse noise is sufficiently suppressed. The system is characterized by a secure bit rate of 2.37~Mbps at 5.6~km and 27.9~kbps at 65.5~km when the fiber dispersion is not compensated. After compensating the fiber dispersion, the QKD distance is extended to 101~km, resulting in a secure key rate of 2.88~kbps. Our results suggest that InGaAs APDs are very well suited to GHz QKD applications.

\end{abstract}

\pacs{03.67.Dd Quantum Cryptography; 85.60.Gz Photo detectors; 85.60.Gw Photodiodes}

\maketitle

Quantum key distribution (QKD)\cite{bennett84,townsend97,gisin02,dusek06} uniquely allows two remote parties (Alice and Bob) to exchange cryptographic keys with verifiable secrecy. Currently, much effort has been focused on increasing the key rate,\cite{bienfang04, gordon05,thew06,diamanti06,takesue07,xu07,namekata07} towards the goal of key exchange being fast enough to offer  information-theoretically secure encryption of high speed data communication. QKD with GHz clock rates\cite{thew06,diamanti06,takesue07} has already been demonstrated in various optical fiber based systems with elevated key rates.  However, due to lack of \textit{practical} detectors at telecom wavelengths, periodically poled LiNbO$_3$ upconversion\cite{thew06, diamanti06} or superconducting nanowire detectors,\cite{takesue07} had to be used.  Upconversion detectors suffer from high background count rates, while superconducting devices require cryogenic cooling to a few K. Neither of these devices are ideal for real-world applications.

\begin{figure}
\begin{center}
\includegraphics[width=8.5cm]{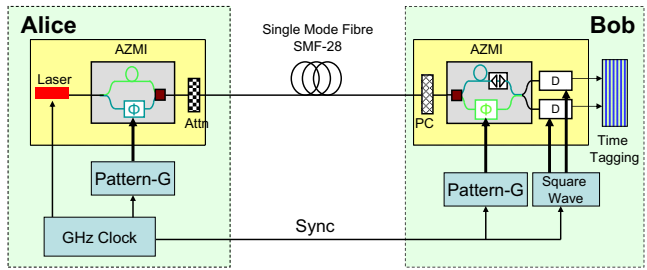}
\caption{
QKD experimental setup. AZMI: asymmetric Mach-Zender interferometer;  Attn: optical intensity attenuator; PC: polarization controller; D: InGaAs APD.}
\end{center}
\end{figure}

Although widely used in the fiber-based MHz QKD systems,\cite{namekata07, gobby04, yoshizawa04,yuan07a} InGaAs avalanche photodiodes (APDs) have been generally regarded as unsuitable for GHz applications due to their severe afterpulsing\cite{ribordy98} problem.
Recently, a self-differencing technique has been shown to suppress the afterpulse noise at high gate frequencies.\cite{yuan07b} It allows detection of weak avalanches previously undetectable under conventional methods, and as a result, the afterpulse noise is sufficiently suppressed to allow APDs to be gated beyond 1~GHz. Moreoever, the self-differencing APD requires no cryogenic cooling, and features a high photon count rate (100~MHz), low dead time ($<10$~ns) and low time jitter (60~ps), suggesting it could be well suited to GHz QKD applications.

In this letter, we demonstrate, for the first time, GHz QKD with GHz-clocked InGaAs APDs. Secure key rates of 2.37 Mbps, 684 kbps and 27.9 kbps have been achieved for fiber lengths of 5.6 km, 25.3 km and 65.5 km respectively. With fiber dispersion compensation, the QKD distance has been extended to 101~km.

The key distribution system is based upon a time division Mach-Zender interferometer using phase modulation,\cite{townsend97,gobby04} as shown in Fig.~1. Photons are generated by a 1.55 $\mu$m pulsed laser operating at 1.036~GHz. The optical pulses are strongly attenuated to the single photon level before leaving the sender's apparatus. Phase modulators in the two interfering arms are used for encoding. A polarization controller is used to recover photon polarizations before Bob's interferometer. Each detected photon is recorded with a unique time stamp using time-tagging electronics. Standard telecom fiber SMF-28 was used in this experiment,  featuring a measured optical attenuation loss of 0.195~dB/km at a wavelength of 1550~nm.

\begin{figure}
\begin{center}
\includegraphics[width=7.5cm]{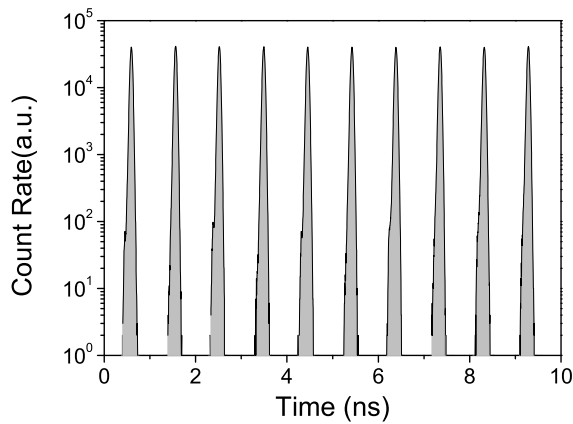}
\caption{Histogram of photon detections with illumination of an average pulse intensity of 0.008 photons at a repetition rate of 1.036 GHz.}
\end{center}
\end{figure}

Self-differencing circuits\cite{yuan07b} were made with co-axial cables of precise lengths so as to match the laser clock frequency.  Our InGaAs APD detectors are cooled typically to --30~$^\circ$C and gated with a square wave of 6.0~V amplitude synchronized to the laser frequency. They have a detection efficiency of $\sim$10\%, dark count probability of $2.9\times10^{-6}$ --- $3.3\times10^{-5}$ and an afterpulse probability of less than 6\%, depending on the bias applied.

Timing performance is of paramount importance for high speed QKD. Figure~2 shows a time-resolved histogram of photon arrivals recorded with a laser illumination of 0.008 photons/pulse.  Each photon peak is very sharp, with a full width at half maximum of 60~ps, suggesting precise timing measurements. Importantly, adjacent peaks are well separated with a 700~ps gap within which no photons were registered at all, due to the nature of gated operation. Such a wide separation not only enables unambiguous bit assignment, but also aids considerable noise rejection.

For experimental simplicity, we choose to implement the standard BB84 protocol\cite{bennett84} with a fixed average optical pulse intensity of $\mu=0.2$. Time-tagged photon detection events are used to precisely measure the raw bit rate ($R_{Raw}$) and the QBER ($e$), from which the secure bit rate after privacy amplification can be determined using the formula\cite{shor00} with a realistic error correction efficiency $f_{EC}=1.10$,\cite{yuan07a}
\begin{equation}
R_{Secure}=\frac{1}{2}R_{Raw}\{1-(1+f_{EC})H(e)\}
\end{equation}
where $\frac{1}{2}$ represents the BB84 protocol efficiency, and  $H(e)=-e{log}_{2}e-(1-e)log_2(1-e)$ is the binary Shannon entropy.\cite{shannon48}

Eqn. (1) would give an \textit{unconditionally} secure\cite{gottesman04} key rate if a truly single photon source were used. The multiphoton pulses, inevitably generated by the attenuated laser in weak pulse QKDs, open up a security threat in which an eavesdropper, Eve, can exploit these extra photons by performing a photon number splitting (PNS) attack.\cite{dusek99,brassard00} However, the PNS attack is currently unfeasible technologically, and more importantly it can be defeated by the recently developed decoy protocol.\cite{lo05, wang05} Our simulations suggest that the decoy protocol would allow an optimized pulse intensity of $\mu=0.52$--0.69, being considerably higher than our choice of $\mu=0.2$, thus ensuring that the QKD demonstrated below can be made unconditionally secure when required in future.  From now on, the PNS attack is excluded from the security analysis.

To permit secure key distribution, the QBER must be lower than a certain threshold, which is $\sim$10\% for our case.  The sources contributing to the QBER can be approximated as
\begin{equation}
e\approx[e_{opt}+\frac{1}{2}P_{a}]+[e_{d}+e_{b}]
\end{equation}
where $e_{opt}$ is the error due to apparatus imperfections, such as interferometer mis-alignment and mis-modulation; $P_{a}$ is the APD afterpulse probability, and the coefficient $\frac{1}{2}$ represents the fact that an afterpulse event has equal probabilities to produce a correct or an erroneous bit. These first two terms are fiber distance independent. $e_{d}$ is the contribution from the detector dark counts. $e_{b}$ is the interclock interference due to pulse broadening, existing only in high speed QKD systems, where an optical pulse may temporally spread into the neighboring clock cycle due to broadening caused by fiber dispersion.
Both $e_d$ and $e_{b}$ are fiber length dependent and significant for long fibers, but are typically negligible at short fiber distances ($\le$25 km). $e_{opt}$ is estimated to be 0.9\%, based on the interferometer visibility (99.4\%) and mis-modulation. $P_a$ depends on the APD bias,\cite{yuan07b} but was found to be below 6\% when the detection efficiency is not greater than 10\%. As shown below, $P_a$ is the dominating error source.

\begin{figure}
\begin{center}
\includegraphics[width=7cm]{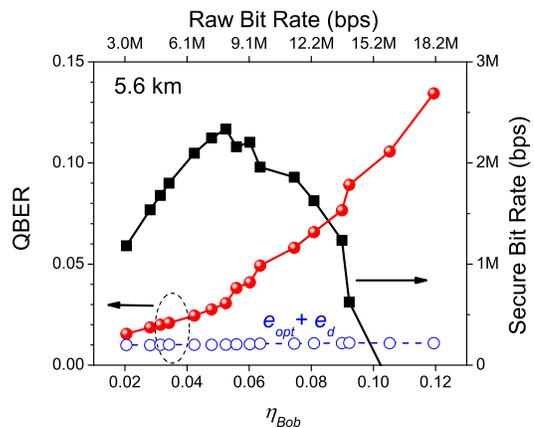}
\caption{
The quantum bit error rate (QBER), and the secure bit rate as a function of Bob's detection efficiency $\eta_{Bob}$ (or raw bit rate) for a fiber link of 5.6~km. Also shown is the QBER arising from the apparatus imperfection ($e_{opt}$)  and detector dark counts ($e_d$).  }
\end{center}
\end{figure}

It is clear from Eqn. (1) that a high raw bit rate and a low QBER are desirable to have a high secure bit rate. However, it is not possible to achieve both simultaneously. Increasing the raw bit rate by raising the detection efficiency always causes an increase in the QBER due to increased dark count rate and afterpulse probability. A trade-off must be made in order to achieve the optimal secure key rate.   To find the optimal operating conditions, we tune the APD DC bias and hence vary the raw bit rate and the QBER.  Figure~3 shows the measured QBER (solid circles) and the secure bit rate (solid square) for a 5.6-km fiber link as a function of Bob's detection efficiency $\eta_{Bob}$ (including Bob's optical loss).  The raw bit rate increases six-fold from 3.1 Mbps to 18.2~Mpbs due to the increased detection efficiency.  The QBER, which is as low as 1.55\% at low detection efficiencies,  increases monotonically, due to the increased noise, and eventually exceeds the security limit of $\sim$10\%. The QBER due to the optical imperfection ($e_{opt}$) and the dark counts ($e_d$) is also plotted, which makes only a small fraction of the total QBER at high detection efficiencies, suggesting that the detector afterpulse may be the dominating error source.
The secure bit rate, calculated using Eqn.~(1),  is highest at an intermediate detection efficiency.
The secure bit rate, which is 1.19 Mbps at low detection efficiency,  doubles to 2.37~Mbps for a detection efficiency between 4\% and 6\%.  Further increasing the detector bias causes a sharp drop in secure bit rate, illustrating that the privacy amplification cost exceeds the gain in raw bit rate. The optimal detector bias was then used for fiber length dependence experiments, corresponding to $\eta_{Bob}$=6.0\%, which gives the optimal secure bit rates over the fiber distances studied.

\begin{figure}
\begin{center}
\includegraphics[width=8cm]{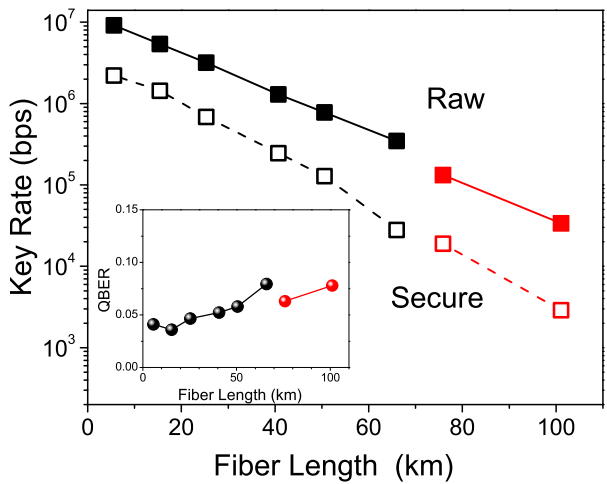}
\caption{
Raw (solid squares) and secure (open squares) bit rates as a function of fiber length. The inset shows the quantum bit error rate (QBER). 75 km and 101~km results were obtained with pre-compensation of fiber chromatic dispersion.}
\end{center}
\end{figure}

We perform QKD experiments over various fiber distances up to 65.5~km. The raw bit rate, as shown in Fig.~4, decreases with fiber length at a rate of 0.24 dB/km. This rate is higher than the measured optical attenuation loss of 0.195~dB/km, because of the additional dispersion loss, as described later. The raw bit rate is 9.16~Mbps at 5.6~km fiber, falling to 348~kbps at 65.5~km.  The measured QBER is shown in the inset of Fig.~4. It remains fairly constant for short fibers below 25~km, but increases with fiber length for longer lengths due to a combination of  the falling photon rate and the increasing deleterious dispersive pulse broadening. $e_b$ is estimated to be 2.1\% for 65.5~km fiber, exceeding the contribution from the detector dark count ($e_d$). Nonetheless, even without fiber dispersion compensation, the QBER at 65.5~km is well below the security limit $\sim$10\%, allowing secure keys to be formed. The calculated secure bit rate follows the raw bit rate, decreasing exponentially with fiber distance. The secure bit rate is determined to be 684~kbps, and 27.9~kbps for 25.3 and 65.5~km respectively.

Secure key exchange over distances much longer than 65~km is not possible without compensation of dispersion of the fiber. The standard fibers used here have a chromatic dispersion of 17~ps/(nm$\cdot$km) at 1550~nm. The chromatic dispersion of the fiber broadens the laser pulse, such that a fraction of photons are dispersed out of the APD detection gate. This causes the raw bit rate to drop more rapidly  (0.24 dB/km) with fiber length than expected from the optical attenuation (0.195 dB/km).  Furthermore, the fiber dispersion also causes temporal overlap of adjacent optical pulses, resulting in an increased interclock interference and error rate. Either effect deteriorates the QKD performance. In fact, the QBER is measured to be 17\% for 75.8~km fiber, having exceeded the security limit, due largely to the contribution of the dispersive broadening ($e_b>10\%$).

We use a fiber Bragg grating device at Alice's side to pre-compensate the fiber dispersion. As shown in the inset of Fig.~4, the QBERs at 75.8~km and 101.1~km have now been reduced to 6.30\% and 7.80\%, respectively, well below the security limit, thus allowing secure key exchange to take place. The secure key rate is determined to be 19.0~kbps and 2.88~kbps for 75.8 and 101.1~km respectively, as shown in Fig.~4.

Finally, we would like to point out that the experiment described here is only a proof-of-principle demonstration. To achieve a complete GHz QKD system, there are still remaining challenges, such as remote synchronization, high speed real-time error correction and privacy amplification, to be addressed.

In summary, we have demonstrated GHz QKD with high bit rate using self-differencing InGaAs APDs over up to 101.1~km standard telecom fiber. The secure bit rate is determined to be 2.37~Mbps for a 5.6~km long fiber, and decreases exponentially to 2.88~kbps for 101.1~km. We conclude that that InGaAs APDs are suitable for long distance GHz QKD systems. These results open the prospect of low cost and practical high bandwidth GHz QKD systems for future applications.

\end{document}